\documentclass[prl,twocolumn,amsmath,showpacs]{revtex4}
\usepackage{graphicx}
\usepackage{amssymb}
\usepackage{citesort}
\usepackage{stmaryrd}

\usepackage{graphicx}

\begin{document}
{\bf Comment on ``Tuning the Magnetic Dimensionality by Charge Ordering
in the Molecular TMTTF Salts''}

Yoshimi {\it et al.} \cite{Yoshimi12a} have attempted to explain the
pressure(P)-dependent behavior of Fabre salts which exhibit charge
order (CO), antiferromagnetic (AFM), and spin-Peierls (SP) phases.
Experiments find two AFM phases \cite{Yu04a,Iwase11a}, AFM$_1$ at
large P and AFM$_2$ at small P.  Yoshimi {\it et al.} suggest that
there also exist two distinct zero-temperature SP phases, SP$_1$ and
SP$_2$.  Here we point out that the occurrence of two distinct SP
phases contradicts experiments \cite{Yu04a,Iwase11a},
and is found in [\onlinecite{Yoshimi12a}] because of unrealistic
model parameters.

Experiments \cite{Yu04a,Iwase11a} emphasize {\it co-operative
  interaction} between the ferroelectric charge order (FCO) and
AFM$_2$ phases.  In the experimental phase diagram
\cite{Yu04a,Iwase11a} $T_{CO}$ and the N\'eel temperature in the
AFM$_2$ phase both decrease with $P$. Thus charge occupancies in the
FCO and AFM$_2$ phases are likely the same. In contrast, P {\it
  increases} \cite{Yu04a,Iwase11a} the SP transition temperature,
indicating that FCO and SP$_2$ phases {\it compete}.  No CO was
detected for $P>$ 0.5 GPa in (TMTTF)$_2$SbF$_6$ \cite{Yu04a,Iwase11a},
in the $P$ region where the SP$_2$ phase occurs at lower temperature.
It is then unlikely that SP$_2$ and FCO coexist at zero temperature.

The hopping parameters used by the authors in their model calculations
are realistic.  Their choice of Coulomb interactions is however
unrealistic.  The onsite Coulomb interaction assumed, $U/t_{a2}$=4, is
too small---in the purely electronic one dimensional model no CO
occurs for this $U$ \cite{Clay03a,Seo06a}.  The assumed intersite
Coulomb interactions $V_b$ = 0 and $V_q=V_a$, are also unrealistic.
Given the lattice geometry (see Fig.~1) it is highly unlikely that
$V_b \ll V_q$, and with large interchain separation $V_q=V_a$ is
equally unrealistic.  $4\alt U\alt 8$ and $V_b$ $\simeq$ $V_q\ll V_a$
is more appropriate.

We repeated the calculations with more realistic $V_a = V$, $V_b = V_q
= 0$, and $4\leq U\leq8$. For these parameters, the intra-dimer charge
structure factor ($C_{-}({\bf q})$ in \cite{Yoshimi12a}) peaks at
several {\bf q} values, indicating comparable energies for both FCO
and the checkerboard pattern CO, in agreement with experiments
\cite{Nakamura03a}.  Peaks in
\begin{figure}[h]
\centerline{\includegraphics[width=3.0in, keepaspectratio=true,trim=0.0in 0.4in 0.0in 0.4in]{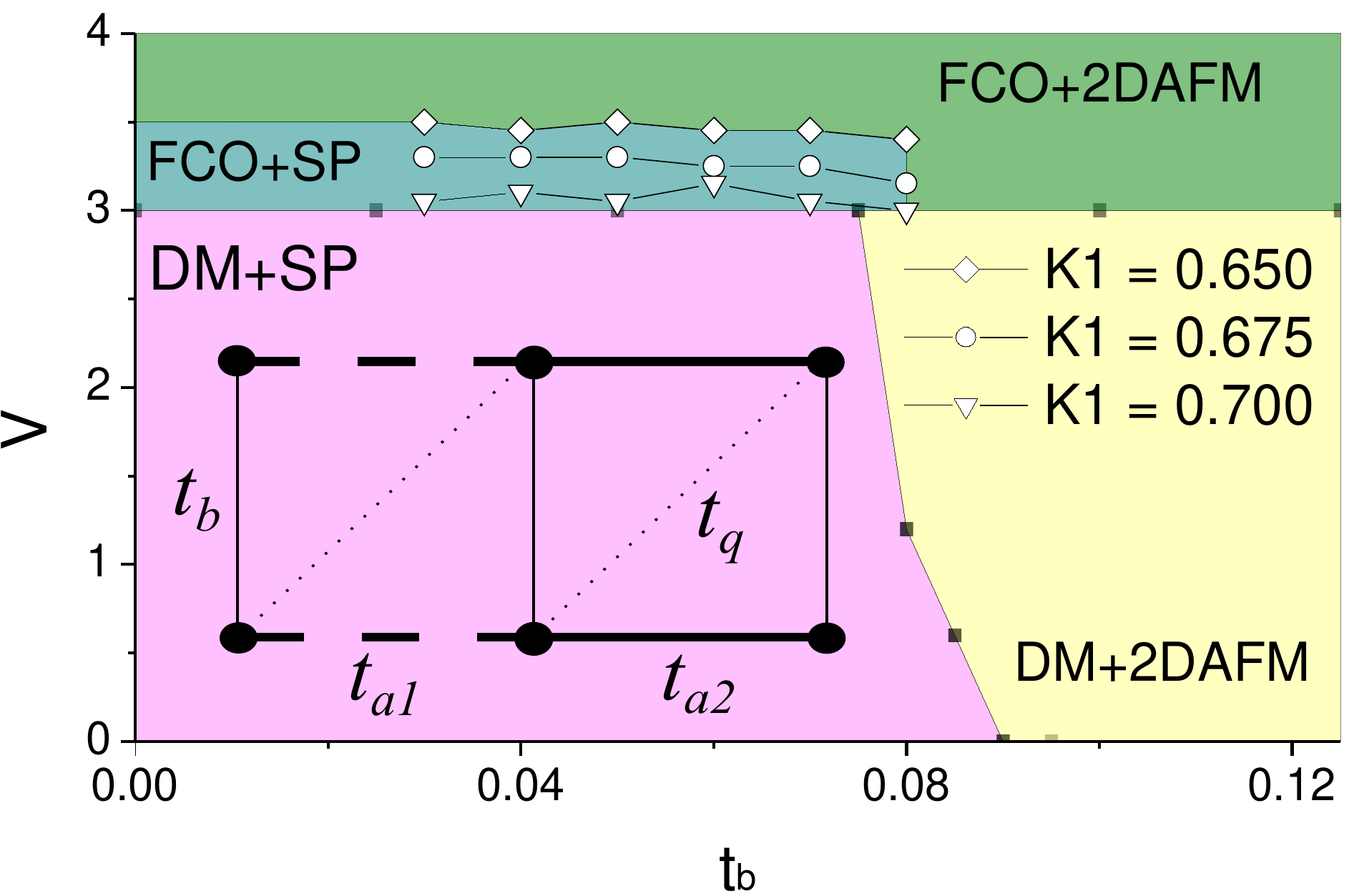}}
\caption{(color online) 8$\times$2 phase diagram for $U=6$, $V_a=V$,
  and $K_2=1$.  The inset shows the lattice structure assumed by
  \cite{Yoshimi12a}.  As $K_1$ increases the size of the FCO+SP phase
  shrinks.  Other points do not significantly change with $K_1$.}
\label{fig1}
\end{figure}
$S_\pm({\bf q})$ remain at the same {\bf q} values as in
Fig.~2 of \cite{Yoshimi12a}.  We conclude that the $V_{ij}$ assumed in
[\onlinecite{Yoshimi12a}] is {\it not} required to explain coexisting
FCO/AFM order in the AFM$_2$ state.

We also repeated (see Fig.~1) the 8$\times$2 calculations with these
parameters.  We have three main observations: (i) for $V_a = V$, $V_b
= V_q = 0$, we find a phase diagram similar to that in
\cite{Yoshimi12a}, but with FCO entering at larger $V$ as expected
\cite{Clay03a,Seo06a}.  The choice $V_q=V$, $V_b=0$ is also {\it not}
required to realize the FCO phase; FCO can be stabilized by
antiferromagnetic superexchange along the $t_b$ bonds; (ii) As $U$
increases the FCO+SP phase narrows; (iii) For both these and the
parameters assumed in \cite{Yoshimi12a}, the width of the FCO+SP phase
is {\it directly proportional to the strength of the inter-site
  electron phonon coupling} (larger $K_1$ gives weaker coupling).
Unconditional transitions in the thermodynamic limit occur in the
limit of 0$^+$ phonon coupling.  Importantly, point (iii) was not
discussed in \cite{Yoshimi12a}, and together with (ii) suggests that
in the thermodynamic limit the FCO+2DAFM and DM+SP phases may share a
common border.

To understand the phase diagram one must consider thermodynamics.  For
large Coulomb interactions the free energy is dominated by spin
excitations.  We have previously shown that the same DM+SP ground
state can have two kinds of soliton spin excitations, (i) with local
CO, or (ii) with uniform charge but local bond distortion
\cite{Clay07a}.  In this picture, to the left of the line bisecting
the SP phase \cite{Yu04a} soliton excitations with local CO dominate
at finite T; to the right occur excitations with uniform site charges.
A unique SP {\it ground state} is expected at all pressures between
AFM$_1$ and AFM$_2$.  We acknowledge support from the Department of
Energy grant DE-FG02-06ER46315.

\medskip
A.~B. Ward$^1$, R.~T. Clay$^1$, and S. Mazumdar$^2$

\hspace{0.2in}\begin{minipage}{3.0in}
$^1$Department of Physics \& Astronomy \\
HPC$^2$ Center for Computational Sciences \\
Mississippi State University \\
Mississippi State, MS 39762-5167 \\
$^2$Department of Physics \\
University of Arizona \\
Tucson, AZ 85721 
\end{minipage}

\end{document}